\documentclass[showpacs,prl,twocolumn,superscriptaddress,amsmath,amssymb,]{revtex4}
\usepackage{color}
\usepackage{graphicx}
\usepackage{epsfig}
\usepackage{dcolumn}
\usepackage{bm}

\begin{document}
\title{Universal quantum fluctuations of a cavity mode driven by a Josephson junction}
\date{\today}
\author{A. D. Armour}
\affiliation{School of Physics and Astronomy, University of Nottingham, Nottingham NG7 2RD, UK}

\author{M. P. Blencowe}
\affiliation{Department of Physics and Astronomy, Dartmouth College, New Hampshire 03755, USA}

\author{E. Brahimi}
\affiliation{Department of Physics and Astronomy, Dartmouth College, New Hampshire 03755, USA}

\author{A. J. Rimberg}
\affiliation{Department of Physics and Astronomy, Dartmouth College, New Hampshire 03755, USA}

\begin{abstract}
We analyze the quantum dynamics of a superconducting cavity coupled to a voltage biased Josephson junction. The cavity is strongly excited at resonances where the voltage energy lost by a Cooper pair traversing the circuit is a multiple of the cavity photon energy. We find that the resonances are accompanied by substantial squeezing of the quantum fluctuations of the cavity over a broad range of parameters and are able to identify regimes where the fluctuations in the system take on universal values.
\end{abstract}
\maketitle

Recent progress in integrating superconducting resonators with Josephson junction devices\,\cite{schoelkopf} and in measuring quantum states in the microwave regime\,\cite{wallraff} has opened up many new ways of using such devices to study the quantum dynamics of nonlinear oscillators\,\cite{qnl,astafiev:07,dc,qnl2,dykman}. Significant attention has been devoted to the idea of using a few-level Josephson `artificial atom' to excite laser-like behavior in a superconducting resonator\,\cite{astafiev:07,you,rodrigues:07a,andre:09,ashhab:09,marthaler:11} and above threshold behavior has now been observed in one such system\,\cite{astafiev:07}. An alternative way of exciting cavity modes which requires neither discrete levels nor an externally applied ac signal, is to harness the Josephson oscillations generated by a dc voltage. Though long studied \,\cite{werthamer:67,stephen,lee,likharev}, such systems are attracting renewed interest given the potential of current experiments to probe the quantum regime in a carefully controlled way.

In the last few years the properties of the photons emitted into a cavity mode by small voltage biased Josephson junctions have been investigated both experimentally\,\cite{hofheinz:11,pashkin:11} and theoretically\,\cite{paduraiu:12,leppa:13}, within the regime where the cavity is close to equilibrium.  However, a very recent experiment used an architecture in which a voltage biased Cooper-pair transistor\,\cite{marthaler:11} is embedded within a superconducting microwave cavity\,\cite{blencowe:12} to achieve a far-from-equilibrium state with a large photon population\,\cite{fei}.

In this Letter we investigate theoretically the quantum dynamics of a model circuit consisting of a voltage biased Josephson junction and a superconducting cavity. We focus on the nonlinear regime where a single cavity mode is strongly excited, deriving a Hamiltonian to describe the behavior close to the family of resonances which occur when the voltage energy lost by Cooper pairs traversing the circuit is an integer multiple of the mode frequency.
The system exhibits quadrature and amplitude squeezing over a broad range of parameters. Surprisingly, there are regimes where the fluctuations take on values that are universal in the sense that they are independent of the system's parameters. We note that a study contemporary with ours\,\cite{ankerhold} investigated a similar system quite independently, but in a very different regime.

\begin{figure}[t]
\centering
{\includegraphics[width=6.0cm]{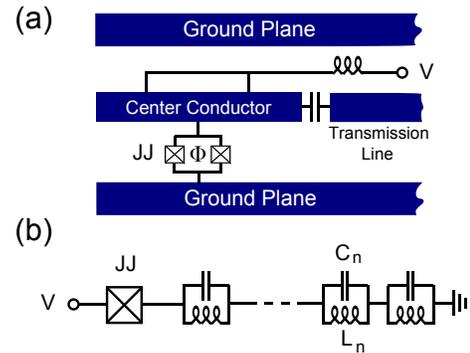}
}
\caption{(a) Schematic diagram and (b) effective circuit model of the system. It consists of two Josephson junctions (JJs) in parallel, forming a single effective junction, in series with a microwave cavity,  modelled as a set of $LC$ oscillators, across which a voltage $V$ is applied. Application of a flux, $\Phi$, to the SQUID loop formed by the junctions allows the effective JJ energy to be tuned. Coupling of the cavity to a transmission line resonator [shown in (a)] leads to dissipation.}
\label{fig:sketch}
\end{figure}

{\it Model System}--- The system we consider is shown schematically in Fig.\ \ref{fig:sketch}a, it consists of two Josephson junctions in parallel mounted on a wire linking the central conductor and ground plane of a superconducting cavity. The parallel combination of two junctions acts as a single effective junction whose Josephson energy can be tuned by applying a suitable flux\,\cite{nakamura}. The dc bias is applied via lines which join the center conductor at points equidistant between the location of the Josephson junctions (sited at the midpoint) and its ends. This geometry (described in detail elsewhere\,\cite{chen:11,blencowe:12,fei}) allows the bias to be applied without affecting the $Q$-factor of cavity modes with voltage nodes at the points where the bias lines join the center conductor. The high $Q$-factors of the modes in the system we consider are crucial: they allow access to the far-from-equilibrium regime where the photon population is large.

A simple effective circuit model for the system is shown in Fig.\ \ref{fig:sketch}b, with the cavity modelled as a set of $LC$ oscillators. The cavity is subject to dissipation arising from couplings between its modes and those of a transmission line resonator.

The Hamiltonian of the effective circuit shown in Fig.\ \ref{fig:sketch}b takes the time-dependent form
\begin{equation}
H=\sum_i\hbar\omega_ia_i^{\dagger}a_i-E_J\cos\left[\omega_dt+\sum_i\Delta_i(a_i+a_i^{\dagger})\right], \label{eq:ham0}
\end{equation}
where $a_i$ is the lowering operator for the $i$-th oscillator with frequency $\omega_i=1/\sqrt{L_iC_i}$, $E_J$ is the effective Josephson junction energy and $\omega_d=2eV/\hbar$ is the frequency associated with the bias voltage $V$. The zero-point displacement of each of the oscillators is given by
$\Delta_i=(e^2Z_i/h)^{1/2}$
where $Z_i=\sqrt{L_i/C_i}$ is the impedance of the $i$-th oscillator\,\cite{blencowe:12}.

Resonances occur when the voltage energy lost by an integer number of Cooper pairs traversing the circuit matches the energy required to create photons in one or more of the cavity modes. Here we will explore resonances of the fundamental cavity mode (with frequency $\omega_0$) which occur when $\omega_d\simeq p\omega_0$, with $p$ an integer and neglect the higher modes in the Hamiltonian \eqref{eq:ham0}.

We analyse the system by moving to a rotating frame defined by $U(t)={\rm e}^{i(\omega_d/p)a^{\dagger}at}$ (dropping the subscript labelling the mode) and derive a time-independent effective Hamiltonian by making a rotating wave approximation (RWA). The RWA should describe the system faithfully when it is very close to resonance, $\omega_0-\omega_d/p=\delta^{(p)}\ll \omega_0,\,\omega_d/p$, provided the couplings $\Delta_0$ and $E_J$ are not too strong. We proceed by expressing the sinusoidal term as exponentials and the Baker-Hausdorff formula\,\cite{gardiner} is used to rewrite the exponentials of $\Delta_0(a+a^{\dagger})$ as a product of exponentials of $a^{\dagger}$ and $a$. This step leads to normal ordering of $a^{\dagger}$, $a$\,\cite{gardiner} and generates a factor of ${\rm e}^{-\Delta_0^2/2}$. We then write out the series expansion of the exponentials, dropping terms with explicit time dependence. Finally, we simplify  using the expansion of the $p$-th Bessel function, $J_{p}(z)=\sum_n (-1)^n(z/2)^{2n+p}/n!(n+p)!$ This results in an effective Hamiltonian
 \begin{eqnarray}
H^{(p)}&=&\hbar\delta^{(p)} a^{\dagger}a-\frac{(-i)^p\tilde{E}_J}{2}:\left[\left(a^{\dagger}\right)^p+(-1)^pa^p\right]\nonumber\\
&&\times\frac{{J}_p(2\Delta_0\sqrt{a^{\dagger}a})}{(a^{\dagger}a)^{p/2}}:, \label{eq:hrwa}
\end{eqnarray}
where $\tilde{E}_J=E_J{\rm e}^{-\Delta_0^2/2}$\,\cite{footnoted} and colons signify normal ordering.

Taking into account weak coupling between the cavity and the modes in the external microwave transmission line, we apply input-output theory\,\cite{gardiner,buks,wallraff} and obtain the Heisenberg equation of motion,
\begin{eqnarray}
\dot{a}&=&-\left(i\delta^{(p)}+\frac{\gamma}{2}\right)a+\sqrt{\gamma}a_{in}\label{eq:heis1}\\
&&+(-i)^{p-1}\frac{\tilde{E}_J\Delta_0}{2\hbar}:\left(\frac{a^{\dagger}}{a}\right)^{(p-1)/2}{J}_{p-1}(2\Delta_0\sqrt{a^{\dagger}a}):\nonumber\\
&&+(-i)^{p-1}\frac{\tilde{E}_J\Delta_0}{2\hbar}:\left(\frac{a}{a^{\dagger}}\right)^{(p+1)/2}{J}_{p+1}(2\Delta_0\sqrt{a^{\dagger}a}):, \nonumber
\end{eqnarray}
where $\gamma$ and the operator $a_{in}$ describe damping and noise, respectively, arising from the coupling to external modes. Assuming zero temperature, the noise operator is described by the correlation functions\,\cite{gardiner}: $\langle a_{in}(t)\rangle=\langle a_{in}^{\dagger}(t)\rangle=0$, $\langle a_{in}(t)a_{in}(t')\rangle=\langle a_{in}^{\dagger}(t)a_{in}^{\dagger}(t')\rangle=\langle a_{in}^{\dagger}(t)a_{in}(t')\rangle=0$ and $\langle a_{in}(t)a_{in}^{\dagger}(t')\rangle=\delta(t-t')/2$.

We can use Eq.\ \eqref{eq:heis1} to obtain an approximate description of the average behavior of the system together with the corresponding fluctuations. We make the replacement $a=\alpha+\delta a$ (and a corresponding one for $a^{\dagger}$),  where $\alpha=\langle a\rangle$ is a complex number  and the operator $\delta a$ describes quantum fluctuations. From  the definition of $\alpha$, we see that the average of the fluctuations must vanish, $\langle \delta a\rangle=0$, and provided they are small we can discard powers of these operators beyond linear order. This amounts to a semiclassical description which also incorporates the zero point fluctuations of the mode\,\cite{gardiner}.

{\it Average dynamics and fluctuations}---The equation of motion for $\alpha$ is obtained by making the replacement $a=\alpha+\delta a$, in Eq.\ \eqref{eq:heis1}, retaining only terms linear order in $\delta a$ and taking the expectation value. Using the definition $\alpha=A{\rm e}^{-i\phi}$ to introduce real variables for amplitude, $A$, and phase, $\phi$, we find
\begin{eqnarray}
\dot{A}&=&-\frac{\gamma}{2}A-\frac{\tilde{E}_J\Delta_0}{2\hbar}\sin[p(\phi-\pi/2)]\nonumber\\
&&\times\left[{J}_{p-1}(2A\Delta_0)+{ J}_{p+1}(2A\Delta_0)\right] \label{eq:mfrwaa}\\
\dot{\phi}&=&\delta^{(p)}-\frac{\tilde{E}_J\Delta_0}{2A\hbar}\cos[p(\phi-\pi/2)]\nonumber\\
&&\times\left[{ J}_{p-1}(2A\Delta_0)-{ J}_{p+1}(2A\Delta_0)\right]. \label{eq:mfrwab}
\end{eqnarray}

The system possesses a rich variety of fixed points\,\cite{werthamer:67} whose locations, $(A_0,\phi_0)$, follow from Eqs.\ \eqref{eq:mfrwaa} and \eqref{eq:mfrwab}. Focusing for simplicity on the cases where the system is on-resonance ($\delta^{(p)}=0$), these points can be divided into three classes. For $p>1$, there is always a fixed point at zero amplitude (though it may not be stable). Beyond this, there are fixed points which owe their existence purely to the presence of dissipation in the system (which we shall refer to as type-I fixed points). These fixed points have phases $\phi_0=\phi^{(p)}_{I}$ with $\cos[p(\phi^{(p)}_I+\pi/2)]=0, \label{eq:sinphi}$
and the amplitudes $A_0=A_I^{(p)}$ are solutions of
\begin{equation*}
A^{(p)}_I=\pm \frac{\tilde{E}_J\Delta_0}{\gamma\hbar}\left(J_{p+1}(2\Delta_0 A^{(p)}_I)+J_{p-1}(2\Delta_0 A^{(p)}_I)\right).
\end{equation*}
Finally, there is a set of points related to the extremal points in the underlying Hamiltonian whose amplitudes, $A_0=A_{II}^{(p)}$, are determined by turning points of Bessel functions, $J_p'(2\Delta_0A_{II}^{(p)})=0$, with phases given by
\begin{equation*}
\sin[p(\phi_{II}^{(p)}-\pi/2)]=-\frac{A_{II}^{(p)}\hbar\gamma}{2\tilde{E}_J\Delta_0 J_{p+1}(2\Delta_0A_{II}^{(p)})}.
\end{equation*}

The equations of motion for the fluctuations  about a given fixed point, $A_0,\phi_0$, take the form
\begin{widetext}
\begin{equation}
\left (\begin{array}{c}\dot{\delta a}\\ \dot{\delta a^{\dagger}}\end{array}\right)=\left (\begin{array}{cc} -i[\delta^{(p)}+\nu_{(p)}(A_{0},\phi_{0})] -\gamma/2 & g_{(p)}(A_{0},\phi_{0})\\ g_{(p)}^*(A_{0},\phi_{0})& +i[\delta^{(p)}+\nu_{(p)}(A_0,\phi_0)] -\gamma/2) \end{array}\right)\left (\begin{array}{c}\delta a\\ \delta a^{\dagger}\end{array}\right) +\sqrt{\gamma}\left (\begin{array}{c}a_{in}\\  a_{in}^{\dagger}\end{array}\right), \label{eq:vec}
\end{equation}
where
\begin{eqnarray*}
\nu_{(p)}(A,\phi)&=&\frac{\tilde{E}_J\Delta_0^2}{\hbar} J_p(2\Delta_0A)\cos[p(\phi-\pi/2)]\\
g_{(p)}(A,\phi)&=&-i\frac{\tilde{E}_J\Delta_0^2}{2\hbar}\left\{J_{p-2}(2\Delta_0A){\rm e}^{i(p-2)(\phi-\pi/2)}+J_{p+2}(2\Delta_0A){\rm e}^{-i(p+2)(\phi-\pi/2)}\right\}.
\end{eqnarray*}
\end{widetext}
The eigenvalues of the matrix in \eqref{eq:vec} determine the stability of the corresponding fixed point; the solution of the coupled equations allows the stationary state fluctuations of the system to be obtained,
\begin{eqnarray*}
\langle \delta a\delta a^{\dagger}+\delta a^{\dagger}\delta a\rangle &=&\frac{\left(\delta^{(p)}+\nu_{(p)}\right)^2+\gamma^2/4}{\left(\delta^{(p)}+\nu_{(p)}\right)^2+\gamma^2/4-|g_{(p)}|^2} \label{eq:var1}\\
\langle \delta a^2\rangle&=&\frac{g_{(p)}}{\gamma+2i\left(\delta^{(p)}+\nu_{(p)}\right)} \langle \delta a\delta a^{\dagger}+\delta a^{\dagger}\delta a\rangle. \label{eq:var2}
\end{eqnarray*}

Fluctuations in the energy of the system are described by the Fano factor $F=(\langle n^2\rangle-\langle n\rangle^2)/\langle n\rangle$. Because the system has a tendency to possess multiple fixed points with the same amplitude, but different phases, amplitude squeezing (characterized by $F<1$) occurs more widely than quadrature squeezing.  For fixed points where $A_0\gg 1$, corrections of order $1/A_0$ can be neglected, leading to 
\begin{equation}
F=\langle\delta a^{\dagger}\delta a+\delta a\delta a^{\dagger}\rangle+{\rm e}^{2i\phi_0}\langle \delta a^2\rangle+ {\rm e}^{-2i\phi_0}\langle (\delta a^{\dagger})^2\rangle. \label{eq:fluc2}
\end{equation}

The Fano factor depends on the particular fixed point the system is at (as we discuss below). The most interesting behavior is seen when the system is at one of the type-II fixed points for which we find (on-resonance)
\begin{equation}
F= \frac{z_{p}J_p(z_{p})}{2\left[z_{p}J_p(z_{p})-pJ_{p+1}(z_{p})\right]}, \label{eq:univ}
\end{equation}
where $z=z_{p}$ is a solution of $J'_p(z)=0$. Remarkably these values depend only on the particular resonance and fixed point involved and are universal in the sense that they are independent of the system's parameters.

{\it One-photon resonance}---We now examine in detail the one-photon resonance ($p=1$) which occurs when $\omega_d=\omega_0$.
Considering the on-resonance case, $\delta^{(1)}=0$, we describe the behavior as $E_J$ is increased from zero\,\cite{footnote2} (this could be achieved in practice by tuning the flux applied to the parallel combination of Josephson junctions\,\cite{nakamura}).

Initially the behavior of the system is controlled by a type-I fixed point with amplitude $A^{(1)}_I=(\tilde{E}_J\Delta_0/\hbar\gamma)[J_0(2\Delta_0 A^{(1)}_I)+J_2(2\Delta_0 A^{(1)}_I)]$, for which the corresponding phase is  $\phi_I^{(1)}=0$.  This point remains the only stable one until $\Delta_0^2\tilde{E}_J/\hbar\gamma=z_1/4J_0(z_1)$, where $z_1\simeq 1.841$ is the first maximum of $J_1(z)$, when a bifurcation occurs. After this bifurcation the fixed point at $A^{(1)}_I$ is unstable and a new pair of type-II stable fixed points emerge with the same amplitudes $A^{(1)}_{II}=z_1/(2\Delta_0)$, but different phases
\begin{equation*}
\phi_{II}^{(1,\pm)}=\pm\arccos\left(\frac{A^{(1)}_{II}\hbar\gamma}{2\tilde{E}_J\Delta_0J_0(2\Delta_0A^{(1)}_{II})}\right).
\end{equation*}

\begin{figure}[t]
\centering
{\includegraphics[width=6.0cm]{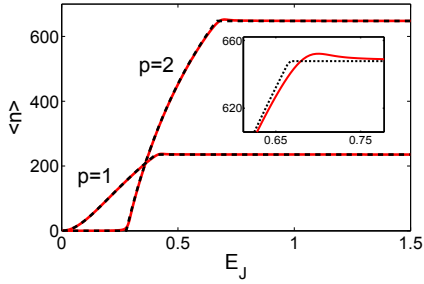}
}
\caption{(Color online) Average energy $\langle n\rangle$ calculated numerically (full curves) compared with stable fixed point amplitudes  (dashed curves) for the $p=1,2$ resonances. The stable fixed point changes from $A_{I}$ to $A_{II}$ at $E_J= 0.405$ ($0.666$) for $p=1(2)$ and the threshold where $A=0$ becomes unstable for $p=2$ is $E_J^c= 0.278$. The inset is a magnified part of the $p=2$ curves.  Adopting units where $\hbar\omega_0=1$, we take $\hbar\gamma= 10^{-3}$, $\Delta_0=0.06$, $\delta^{(p)}=0$, values which are used throughout.}
\label{fig:nav}
\end{figure}

Figure \ref{fig:nav} shows how the average energy of the steady state, $\langle n\rangle=\langle a^{\dagger}a\rangle$, evolves as a function of $E_J$. The results are compared with a numerical solution of the Lindblad master equation equivalent to Eq.\ \eqref{eq:heis1}, which provides a check on the validity of the analytical approach\,\cite{qutip}.
The semiclassical prediction (given by the square of the relevant stable fixed point amplitude) describes the behavior extremely well except close to the bifurcation. The energy evolves smoothly from lower to higher energies as $E_J$  is increased.

The absence of a threshold is easily understood by expanding Eq.\ \eqref{eq:hrwa} for $p=1$ to lowest order in $\Delta_0$, the resulting Hamiltonian, $H^{(1)}\simeq\hbar\delta^{(1)} a^{\dagger}a+i(E_J\Delta_0/2)(a^{\dagger}-a)$
describes a (linearly) driven harmonic oscillator. This approximate Hamiltonian accounts for the initial quadratic growth of $\langle n\rangle$ seen in Fig.\ \ref{fig:nav}, but is inadequate when nonlinear effects become important. Above the bifurcation $\langle n\rangle$ is insensitive to both $E_J$ and $\gamma$, but varies as $1/\Delta^2_0$.

\begin{figure}[t]
\centering
{\includegraphics[width=6.0cm]{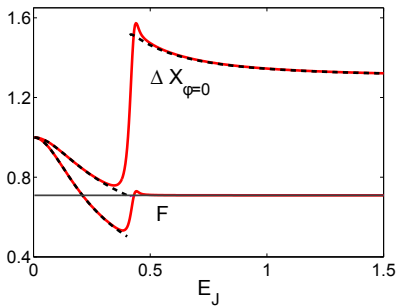}
}
\caption{(Color online) Fluctuations in the energy, $F$, (lower full curve) and quadrature (upper full curve), $\Delta X_{\phi=0}$, calculated numerically as a function of $E_J$ for $p=1$. Analytic results for the fixed points $A^{(1)}_I$ (for $E_J<0.405$) and $A^{(1)}_{II}$ (for $E_J\ge 0.405$) are shown as dashed curves. The horizontal line is the value of $F$ given by Eq.\ \eqref{eq:univ}.  }
\label{fig:unc}
\end{figure}

The fluctuations of the cavity mode are shown in Fig.\ \ref{fig:unc}. Amplitude squeezing occurs across the whole parameter regime studied and quadrature squeezing below the bifurcation between the type-I and type-II fixed points. Starting at the type-I fixed point and  moving towards the bifurcation, the linearized theory predicts $F\rightarrow0.5$. For $p=1$ amplitude squeezing coincides with quadrature squeezing at the type-I fixed point with $\Delta X_{\phi=0}^2=F$, where we define the quadrature $X_{\phi}=a{\rm e}^{-i\phi}+a^{\dagger}{\rm e}^{i\phi}$.
Above the bifurcation $F$ saturates rapidly to the universal value $0.7092$ given by Eq.\ \eqref{eq:univ} for $p=1$.

{\it Two-photon resonance}---Next we turn to the behavior of the system at the two-photon ($p=2$) resonance. $A=0$ is now a fixed point and consequently the system displays a threshold; significant occupation of the cavity only occurs when this fixed point is unstable. For weak couplings and low photon numbers we can again expand the Hamiltonian to lowest order in $\Delta_0$; in this case $H^{(2)}\simeq \hbar\delta^{(2)}a^{\dagger}a+E_J(\Delta_0/2)^2(aa+a^{\dagger}a^{\dagger})$.
This limiting form of the Hamiltonian is that of a degenerate parametric amplifier (DPA)\,\cite{parametric}  and the two systems behave in the same way in the below-threshold regime\,\cite{paduraiu:12}.

\begin{figure}[t]
\centering
{\includegraphics[width=6.0cm]{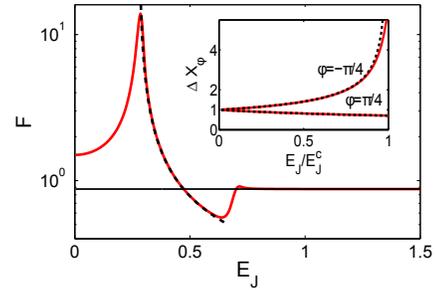}
}
\caption{(Color online) Fluctuations at the $p=2$ resonance. The main plot shows $F$, numerics (full curve) are compared with analytic results when $A_0>0$: the dashed curve is for the $A_I$ fixed points and the horizontal line,  given by Eq.\ \eqref{eq:univ}, describes the $A_{II}$ fixed points. Threshold is $E_J^c=0.278$ and the bifurcation between $A_I$ and $A_{II}$ fixed points is at $E_J=0.666$.  The inset compares analytic (dashed curves) and numerical (full curves) calculations of $\Delta X_{\phi=\pm\pi/4}$ below threshold.}
\label{fig:tpn}
\end{figure}

On-resonance ($\delta^{(2)}=0$), the $A=0$ fixed point is stable for $\tilde{E}_J\Delta_0^2/\hbar \gamma<1$, leading to a region of very small occupation numbers. Above this threshold a pair of type-I fixed points is stable, they have the same amplitude
\begin{equation}
A_I^{(2)}=\frac{\tilde{E}_J\Delta_0}{\hbar}\left[J_0(2\Delta_0A_I^{(2)})+J_3(2\Delta_0A_I^{(2)})\right],
\end{equation}
but different phases $\phi_I^{(2,\pm)}=\pi/4,\pi/4+\pi$. These fixed points in turn become unstable with a bifurcation at $\Delta_0^2\tilde{E}_J/\hbar\gamma=z_2/4J_1(z_2)$, where $z_2\simeq 3.054$ is the first maximum of $J_2(z)$, and a new set of type-II stable fixed points emerge with amplitude   $A^{(2)}_{II}=z_2/(2\Delta_0)$.

The average energy of the system for $p=2$ is shown as a function of $E_J$ in Fig.\ \ref{fig:nav}. There is a clear threshold, then when the system reaches the type-II fixed points 
the energy becomes independent of $E_J$.
The fluctuations of the cavity mode are shown in Fig.\ \ref{fig:tpn}. Below threshold
the system behaves like the DPA\,\cite{parametric} displaying squeezing of $X_{\phi=\pi/4}$.  The linear theory predicts $\Delta X^2_{\phi=\pi/4}\rightarrow0.5$ as the threshold is approached from  below whilst the uncertainty in the conjugate quadrature, $\Delta X_{\phi=-\pi/4}$, diverges.
The threshold is accompanied by a peak in $F$, which then drops abruptly and the linear theory again gives $F\rightarrow 0.5$ at the bifurcation between the type-I and type-II fixed points. Above the second bifurcation $F$ goes to the universal value $0.8753$ predicted by Eq.\ \eqref{eq:univ}.

{\it Conclusions}---We derived an effective Hamiltonian describing an experimentally accessible Josephson junction-cavity system close to resonances which occur when Cooper pairs crossing the junction excite photons in a cavity mode. The system displays amplitude and quadrature squeezing for a wide ranges of parameters. Furthermore, the amplitude fluctuations of the cavity mode can take universal values which
are independent of the system's parameters. Our work provides a starting point for several future studies.  The RWA Hamiltonian can be used to investigate properties of the system beyond the linear fluctuations\,\cite{dykman,spectrum,drummond}. It could also be extended to include two or more cavity modes, the additional modes might provide another source of dissipation and for appropriate bias voltages two modes would be excited\,\cite{hofheinz:11}.

\begin{acknowledgments}
We thank J. Ankerhold for helpful discussions. ADA was supported by EPSRC (UK), Grant No. EP/I017828. MPB and AJR were supported by the NSF (grants DMR-1104790 and DMR-1104821) and by AFOSR/DARPA agreement FA8750-12-2-0339.
\end{acknowledgments}


\begin{thebibliography}{99}
\bibitem{footnoted} Note that we expect  $\Delta_0\ll 1$ in practice (see e.g.\ \cite{blencowe:12}) and hence we anticipate that $\tilde{E}_J$ and $E_J$ will be the same to a very good approximation for experimentally accessible systems.
\bibitem{footnote2}It is straightforward to map out the equivalent behavior as either  $\Delta_0$ or $\gamma$ is varied, but these values would most likely be fixed in a given experiment.
\end{thebibliography}
\end{document}